# Generation and manipulation of entangled photons in a domain-engineered lithium niobate waveguide

Yang Ming,[1,3] Ai-hong Tan,[2] Zi-jian Wu,[1,3] Zhao-xian Chen,[1,3] Fei Xu[1,3,*] and Yan-qing Lu[1,3,*]

[1]*National Laboratory of Solid State Microstructures and College of Engineering and Applied Sciences, Nanjing University, Nanjing 210093, China*
[2]*Laboratory for Quantum Information, China Jiliang University, Hangzhou 310018, China*
[3]*National Center of Microstructures and Quantum Manipulation, Nanjing University, Nanjing 210093, China*
**Corresponding author:* yqlu@nju.edu.cn , feixu@nju.edu.cn*

We propose to integrate the electro-optic tuning function into polarization-entangled photon pair generation process in a periodically poled lithium niobate (PPLN). Due to the versatility of PPLN, both the spontaneously parametric down conversion and electro-optic polarization rotation effects could be realized simultaneously. Orthogonally-polarized and parallel-polarized photon pairs thus are instantly switchable by tuning the applied field. The characteristics of the source are investigated showing adjustable bandwidths and high entanglement degrees. Moreover, other kinds of reconfigurable entanglement are also achievable based on suitable domain-design. We believe the domain engineering is a very promising solution for next generation function-integrated quantum circuits.

*OCIS codes:* (270.0270) Quantum optics; (160.2100) Electrooptical materials; (160.3730) Lithium niobate.

## 1. INTRODUCTION

Entangled photon pairs are crucial physical resource for quantum information science and technology. Generating, manipulating and detecting entangled photons constitute the basic processes of quantum information applications. To generate entangled photons, spontaneous parametric down conversion (SPDC) in $\chi^{(2)}$ nonlinear crystals, such as lithium niobate (LN) [1] or beta barium borate (BBO) [2], is one of the most powerful tools to transform a single photon into an entangled photon pair. To manipulate entangled photons, external electro-optic (EO) tuning is an effective way so that the phase and polarization of photon pairs could be instantly reconfigured [3]. However, for typical ferroelectric nonlinear optical materials, such as LN, they combine both the SPDC and EO tuning capabilities together, function integration thus is feasible, which makes the source reconfigurable. Actually, LN is also a good platform to integrate other functions together, such as wave-guiding and beam splitting. Even the efficient infrared photon detection could be realized based on LN [4]. Therefore LN has been treated as a potential candidate for integrated quantum information processing.

What's more, the ferroelectric domain structure of LN is engineerable, providing more attractive nonlinear [5], EO [6] and negative permittivity [7] properties in a periodically poled lithium niobate (PPLN). Highly efficient entangled source based on PPLN waveguide has been demonstrated [5]. Recently, on-chip spatial control [8] and lensless ghost imaging [9] of entangled photons also have been realized through two-dimensional domain engineering in LN substrates. It is quite interesting if the unique EO polarization rotation effect also could be implemented into a PPLN or a more complex domain-engineered LN. Versatile functions including enchantment source, EO quantum logic gates [10], instantly beam manipulation and efficient photon detection all could be integrated together toward future practical large scale quantum circuit integration.

In this letter, we propose an approach to combine EO and SPDC processes for generating and manipulating polarization-entangled state in a PPLN waveguide with suitable designed artificial structures. Based on the approach, a reconfigurable polarization entangled photon pair source is established. When a voltage is applied, the source produces a pair of entangled photons contains the same polarization, with either o- or e- polarization state. In contrast, if the voltage is turned off, the entangled photons bear orthogonal polarization states, *i.e.*, one photon is o-polarized while another one has e-polarization. Since we may control the polarization state intentionally, the formations of entangled states thus could be regulated accordingly, which possess considerable potential in modulation of multi-photon entangled states.

## 2. STRUCTURE DESIGN OF DUAL PERIODICALLY POLED LITHIUM NIOBATE

We consider the processes of photon interaction in a $z$-cut, $x$-propagating titanium in-diffused PPLN waveguide. In this kind of nonlinear periodic structure, quasi-phase-matching (QPM) is a powerful solution to provide effective nonlinear interaction through modulation of nonlinear coefficients, no matter for SPDC [5] or even EO [6] processes. Moreover, the domain structures could be engineered to support multiple photon interaction processes, leading to integrated photonic devices. Here, as an example, a simple dual-periodic PPLN waveguide is designed to realize the tunable photon entanglement. As shown in Fig. 1, the domains exhibit twice-modulation upon a periodic grating so that they may supply sufficient reciprocal vectors for several SPDC and EO processes. As a consequence, if a suitable domain design is given, the EO polarization rotation [6] could realize simultaneously together with two SPDC processes. The corresponding SPDC processes are expressed as $o_p \to o_s + e_i$, $o_p \to e_s + o_i$, where $p$, $s$, and $i$ represent the pump, signal, and idler waves, respectively. That composes a type-II-like entangled photon pair with polarization states of $(o_s, e_i)$ or $(e_s, o_i)$. Moreover, if a suitable field is applied, the EO process takes place for the signal wave, namely $o_s \to e_s$ ($e_s \to o_s$). Thus the entangled pair is modulated to be type-I-like with the same polarization states, i.e., $(o_s, o_i)$ or $(e_s, e_i)$.

To compensate QPM conditions of the three interaction processes, the inverted domains of PPLN play a critical role. The sign of nonlinear coefficient $\chi^{(2)}$ and EO coefficient $\gamma$ change periodically. The modulation functions are expressed as $d(x) = d_{31}f_1(x)f_2(x)$ and $\gamma(x) = \gamma_{51}f_1(x)f_2(x)$ [11], where $f(x) = \text{sign}[\cos(2\pi x/\Lambda)] = \sum_m G_m e^{ik_m x}$. In the equations, $d$ is the substitute of nonlinear coefficient $\chi^{(2)}$ with the relationship $d = \chi^{(2)}/2$. The effective component of $d$ and $\gamma$ in our situation are $d_{31}$ and $\gamma_{51}$, respectively. Thus the reciprocal vectors are given as $k_m = 2m\pi/\Lambda$, with their corresponding Fourier coefficients as $G_m = (2/m\pi)\sin(m\pi/2)$. In all, the expansion formulation of the coefficients are obtained as [12]

$$d(x) = d_{31} \sum_{m,n} G_{m,n} e^{iK_{m,n}x}$$
$$\gamma(x) = \gamma_{51} \sum_{m,n} G_{m,n} e^{iK_{m,n}x} \qquad (1)$$

with

$$G_{m,n} = \frac{4}{mn\pi^2} \sin(\frac{m\pi}{2}) \sin(\frac{n\pi}{2})$$

$$K_{m,n} = \frac{2m\pi}{\Lambda_1} + \frac{2n\pi}{\Lambda_2}$$

where $\Lambda_1$ and $\Lambda_2$ represents the corresponding two modulation periods. Based on these analyses, the equations of QPM conditions could be expressed as

$$\Delta\beta_1 = \beta_{p,o} - \beta_{s,o} - \beta_{i,e} = K_{m_1,n_1}$$
$$\Delta\beta_2 = \beta_{p,o} - \beta_{s,e} - \beta_{i,o} = K_{m_2,n_2} \qquad (2)$$
$$\Delta\beta_3 = \beta_{s,o} - \beta_{s,e} = K_{m_3,n_3}$$

In these equations, $\beta_{j,\sigma}$ ($j = p, s, i; \sigma = o, e$) refers to the propagation constant of the corresponding waveguide mode (fundamental modes are assumed), which is obtained based on the Hermite-Gauss formulations [13]. $\Delta\beta_1$ and $\Delta\beta_2$ correspond to two SPDC processes, while $\Delta\beta_3$ corresponds to the EO interaction. $K_{m_i,n_i}$ (i=1, 2, 3) are the required reciprocal vectors to compensate the phase mismatch.

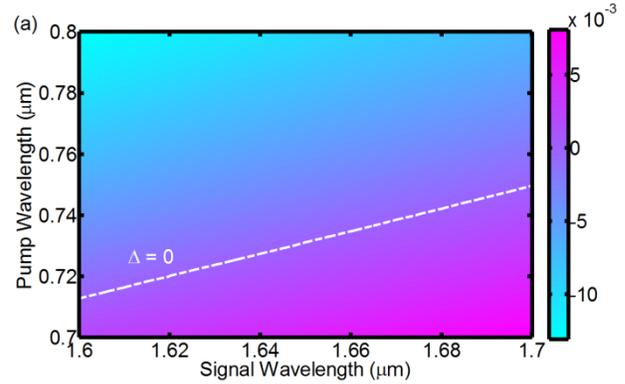

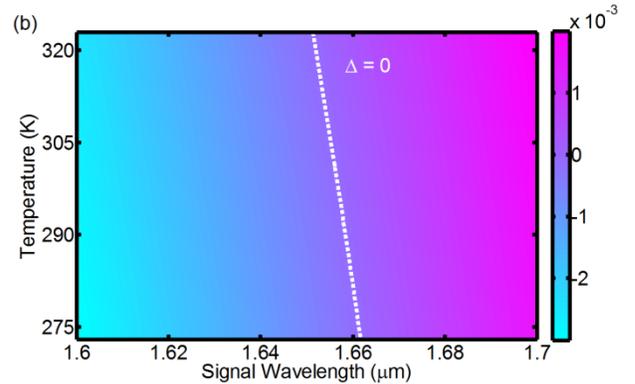

Fig. 2. Illustrations of the representations of phase mismatches, i.e., $\Delta(\lambda)$. When $\Delta(\lambda) = 0$, three phase matching conditions could be satisfied simultaneously. The white dashed line marks $\Delta(\lambda) = 0$. (a) Values of $\Delta(\lambda)$ change with the signal wavelengths and the pump wavelengths. (b) Values of $\Delta(\lambda)$ changes with the signal wavelengths and the operation temperature. From the $\Delta(\lambda) = 0$ lines, suitable phase matching conditions could be chosen.

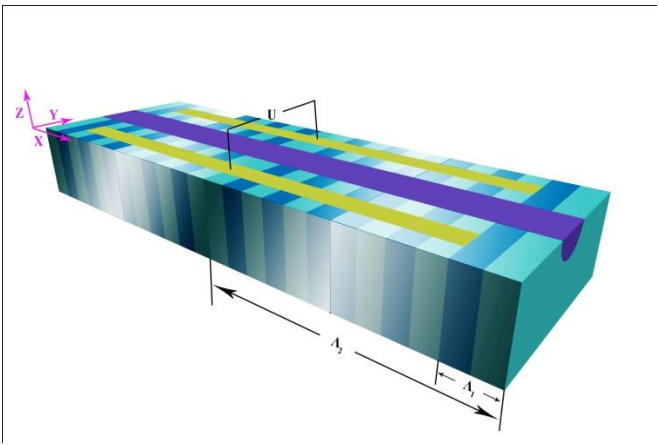

Fig. 1. Schematic of the dual-PPLN waveguide. The light and dark blue portions represent the positive and negative domains of the PPLN, respectively; while the purple portion is the core of the waveguide. The two golden strips correspond to the electrodes. $\Lambda_1$ and $\Lambda_2$ are the two periods of the structure.

Through detailed calculations, we find an appropriate set of solutions for Eq. (2). The width and depth of the waveguide core are both set at 10 μm, and the maximum of index difference is set at 0.003. The pump, signal and idler wavelength are set at 0.7335 μm, 1.6568 μm and 1.3162 μm, respectively, while the operation temperature is set at 25 °C. The corresponding values for the ($m$, $n$) series are {($m_1$, $n_1$) = (3, 1), ($m_2$, $n_2$) = (3, -1), ($m_3$, $n_3$) = (1, 1)}. The twice-modulation periods are designed at $\Lambda_1$ = 25.84 μm and $\Lambda_2$ = 154.96 μm, respectively. The ratio of these two periods is $\Lambda_2/\Lambda_1$ = 6 and the duty cycle is 0.5, as is shown in Fig. 1. Moreover, the phase matching conditions could be adjusted by the pump wavelength and the operation temperature, which are shown in Fig. 2(a) and Fig. 2(b). The quantity $\Delta(\lambda)$ displayed in the figures is obtained as follows: we make $K_{3,1}$ and $K_{3,-1}$ equal to $\Delta\beta_1$ and $\Delta\beta_2$, respectively, then a set of $\Lambda_1$ and $\Lambda_2$ is calculated. $\Delta(\lambda)$ = $\Delta\beta_3(\lambda) - G_{1,1}$ thus are obtained. As long as $\Delta(\lambda) = 0$, which represented by the white dash lines, $\Delta\beta_1$, $\Delta\beta_2$ and $\Delta\beta_3$ could be compensated simultaneously.

## 3. QUANTUM ANALYSIS OF ELECTRO-OPTICALLY TUNABLE ENTANGLEMENT

To give a quantum mechanics description of our tunable entangled source, we derive the state vector of entangled photons through the effective Hamiltonian. The total interaction Hamiltonian consists of two parts, which is expressed as $H_1$ = $H_{SPDC}$ + $H_{EO}$. $H_{SPDC}$ corresponds to the SPDC processes, while $H_{EO}$ refers to the EO interaction. Both of them arise from polarization, namely, $P_{NL}$ and $P_{EO}$. In detailed treatments, the pump field is usually treated as an undepleted classical wave. The signal and idler fields are quantized and represented by field operators. Thus the formulations of the pump, signal and idler fields could be written as

$$E_p^{(+)} = E_{po}\vartheta_{po}(\vec{r})e^{i(\beta_{po}x-\omega_p t)}$$

$$E_s^{(-)} = i\sum_\sigma \int d\omega_s \sqrt{\frac{\hbar\omega_s}{2\varepsilon_0 n_{s\sigma}^2 N_{s\sigma}}}\vartheta_{s\sigma}(\vec{r})a_{s\sigma}^\dagger e^{-i(\beta_{s\sigma}x-\omega_s t)} \quad (3)$$

$$E_i^{(-)} = i\sum_\sigma \int d\omega_i \sqrt{\frac{\hbar\omega_i}{2\varepsilon_0 n_{i\sigma}^2 N_{i\sigma}}}\vartheta_{i\sigma}(\vec{r})a_{i\sigma}^\dagger e^{-i(\beta_{i\sigma}x-\omega_i t)}$$

where σ represents the polarization state o or e, and $n_{j\sigma}$ (j=s, i) is the corresponding refractive index. $N_{s\sigma}$ and $N_{i\sigma}$ are normalization parameters. $\vartheta_{po}(r)$, $\vartheta_{so}(r)$ and $\vartheta_{io}(r)$ correspond to the transverse mode profiles of pump, signal and idler field, respectively.

For SPDC process, using the rotating wave approximation, $H_{SPDC}$ is derived as

$$H_{SPDC} = \frac{1}{2}\int_V d^3 r P_{NL}\cdot E = \varepsilon_0 \int_V d^3 r d(x)E_p^{(+)}E_s^{(-)}E_i^{(-)} + H.C.$$

$$= -\frac{\hbar E_{po}}{2}\sum_{m,n} d_{31}G_{m,n}[\int_{-L}^0 dx e^{i(\beta_{po}-\beta_{so}-\beta_{ie}+K_{m,n})x}$$

$$\times \iint d\omega_s d\omega_i F_{oe}\sqrt{\frac{\omega_s\omega_i}{n_{so}^2 n_{ie}^2 N_{so}N_{ie}}}a_{so}^\dagger a_{ie}^\dagger e^{-i(\omega_p-\omega_s-\omega_i)t} \quad (4)$$

$$+\int_{-L}^0 dx e^{i(\beta_{po}-\beta_{se}-\beta_{io}+K_{m,n})x}\iint d\omega_s d\omega_i F_{eo}\sqrt{\frac{\omega_s\omega_i}{n_{se}^2 n_{io}^2 N_{se}N_{io}}}$$

$$\times a_{se}^\dagger a_{io}^\dagger e^{-i(\omega_p-\omega_s-\omega_i)t} + H.C.]$$

with

$$F_{oe} = \iint dydz\vartheta_{po}(\vec{r})\vartheta_{so}(\vec{r})\vartheta_{ie}(\vec{r})$$
$$F_{eo} = \iint dydz\vartheta_{po}(\vec{r})\vartheta_{se}(\vec{r})\vartheta_{io}(\vec{r})$$

To obtain $H_{EO}$, we start with the polarization $P_{EO}$. It could be expressed as $P_j = -\sum_k \varepsilon_0 \gamma_{51} E_a n_o^2 n_e^2 \Delta\varepsilon_{jk} E_k$ with $\Delta\varepsilon_{jk} = \begin{cases} 1, & jk = (23, 32) \\ 0, & else \end{cases}$

where $E_a$ is the applied electric field. Thus we have

$$H_{EO} = \frac{\hbar\varepsilon_0}{4}\sum_{m,n}\gamma_{51}E_a G_{m,n}[\int_{-L}^0 dx e^{i(\beta_{s'e}-\beta_{so}+K_{m,n})x}$$

$$\times \iint d\omega_s d\omega_{s'} F_{EO}\sqrt{\frac{\omega_s \omega_{s'} n_{so}^2 n_{s'e}^2}{N_{so}N_{s'e}}}a_{so}^\dagger a_{s'e} e^{-i(\omega_s-\omega_{s'})t} \quad (5)$$

$$+\int_{-L}^0 dx e^{i(\beta_{so}-\beta_{s'e}+K_{m,n})x}\iint d\omega_s d\omega_{s'} F_{EO}\sqrt{\frac{\omega_s \omega_{s'} n_{so}^2 n_{s'e}^2}{N_{so}N_{s'e}}}$$

$$\times a_{s'e}^\dagger a_{so} e^{-i(\omega_{s'}-\omega_s)t}]$$

with

$$F_{EO} = \iint dydz\vartheta_{so}(\vec{r})\vartheta_{se}(\vec{r})$$

The first and second items represent the transformation of the signal photon from e-polarization to o-polarization and vice versa. When the applied electric field $E_a$ is turned on, $H_{EO}$ combines with $H_{SPDC}$ to make the entangled state transform from parallel-polarization into orthogonal-polarization. The entangled state vector is derived as $|\Psi(t)\rangle = \exp[(-i/\hbar)\int dt(H_{SPDC} + H_{EO}(E_a))]|0\rangle$. Detailed treatments thus could be divided into two cases:

i) when there is no applied voltage, we expand the evolution operator to the first-order perturbation, so the state vector is expressed as

$$|\Psi\rangle = [1-\frac{i}{\hbar}\int_{-\infty}^\infty dt H_{SPDC}(t)]|0\rangle$$

$$= |0\rangle + \frac{iE_{po}}{2}\sum_{m,n} d_{31}G_{m,n}\iint d\omega_s d\omega_i [F_{oe}\sqrt{\frac{\omega_s\omega_i}{n_{so}^2 n_{ie}^2 N_{so}N_{ie}}}$$

$$\times \int_{-L}^0 dx e^{i(\beta_{po}-\beta_{so}-\beta_{ie}+K_{m,n})x}\int_{-\infty}^\infty dt e^{-i(\omega_p-\omega_s-\omega_i)t}a_{so}^\dagger a_{ie}^\dagger \quad (6)$$

$$+F_{eo}\sqrt{\frac{\omega_s\omega_i}{n_{se}^2 n_{io}^2 N_{se}N_{io}}}\int_{-L}^0 dx e^{i(\beta_{po}-\beta_{se}-\beta_{io}+K_{m,n})x}$$

$$\times \int_{-\infty}^\infty dt e^{-i(\omega_p-\omega_s-\omega_i)t}a_{se}^\dagger a_{io}^\dagger]|0\rangle$$

In the equation, we have $\int dt e^{-i(\omega_p-\omega_s-\omega_i)t} = 2\pi\delta(\omega_p-\omega_s-\omega_i)$ and $\int dx e^{i\Delta\beta x} = h(L\Delta\beta)$ with the formation $h(x) = e^{(-ix/2)}sinc(x/2)$. Moreover, the actual SPDC process doesn't merely arise at the perfect phase-matching frequencies $\Omega_s$ and $\Omega_i$. As the natural bandwidth is ν, we set $\omega_s = \Omega_s + \nu$ and $\omega_i = \Omega_i - \nu$, then the state vector is simplified as

$$|\Psi\rangle = |0\rangle + \int d\nu P_{oe}h(L\Delta\beta_{ooe})a_{so}^\dagger a_{ie}^\dagger |0\rangle$$
$$+\int d\nu P_{eo}h(L\Delta\beta_{oeo})a_{se}^\dagger a_{io}^\dagger |0\rangle \quad (7)$$

with

$$P_{oe} = i\pi L E_{po} d_{31} G_{3,1} F_{oe} \sqrt{\frac{\Omega_s \Omega_i}{n_{so}^2 n_{ie}^2 N_{so} N_{ie}}}$$

$$P_{eo} = i\pi L E_{po} d_{31} G_{3,-1} F_{eo} \sqrt{\frac{\Omega_s \Omega_i}{n_{se}^2 n_{io}^2 N_{se} N_{io}}}$$

As $|v| \ll \Omega_s$, $\Omega_i$, we have substituted $\omega_s(\omega_i)$ by $\Omega_s$ ($\Omega_i$) in $P_{oe}$ and $P_{eo}$.

ii) For the second case, the voltage $E_a$ is applied. The evolution operator should be expanded to the second-order term. The formulation of the state vector could be expressed as

$$|\Psi\rangle = [1 + (-\frac{i}{\hbar})^2 \int_{-\infty}^{\infty} dt_2 \int_{-\infty}^{\infty} dt_1 T(H_{EO}(t_2) H_{SPDC}(t_1))]|0\rangle$$

$$= |0\rangle + \frac{\varepsilon_0 d_{31} \gamma_{51} E_{po} E_a L^2}{8} \{\iint d\omega_s d\omega_{s'} \int_{-\infty}^{\infty} dt_2 e^{-i(\omega_s - \omega_{s'})t_2} F_{EO}$$

$$\times [G_{1,1} h(L\Delta\beta_{eo}) \sqrt{\frac{\omega_s \omega_{s'} n_{so}^2 n_{s'e}^2}{N_{so} N_{s'e}}} a_{so}^\dagger a_{s'e} + G_{1,1} h(L\Delta\beta_{oe})$$

$$\times \sqrt{\frac{\omega_s \omega_{s'} n_{so}^2 n_{s'e}^2}{N_{so} N_{s'e}}} a_{s'e}^\dagger a_{so}]\} \{\iint d\omega_s d\omega_i \int_{-\infty}^{\infty} dt_1 e^{-i(\omega_p - \omega_s - \omega_i)t_1}$$

$$\times [G_{3,1} h(L\Delta\beta_{ooe}) F_{oe} \sqrt{\frac{\omega_s \omega_i}{n_{so}^2 n_{ie}^2 N_{so} N_{ie}}} a_{so}^\dagger a_{ie}^\dagger$$

$$+ G_{3,-1} h(L\Delta\beta_{oeo}) F_{eo} \sqrt{\frac{\omega_s \omega_i}{n_{se}^2 n_{io}^2 N_{se} N_{io}}} a_{se}^\dagger a_{io}^\dagger]\} |0\rangle \quad (8)$$

The functions $h(L\Delta\beta_{ooe})$ and $h(L\Delta\beta_{oeo})$ are spatial integration for SPDC, while $h(L\Delta\beta_{oe})$ and $h(L\Delta\beta_{eo})$ correspond to the EO processes. There are four product terms of effective operators, including $a_{so}^\dagger a_{se} a_{se}^\dagger a_{io}^\dagger$, $a_{se}^\dagger a_{so} a_{so}^\dagger a_{ie}^\dagger$, $a_{so}^\dagger a_{se} a_{so}^\dagger a_{ie}^\dagger$, $a_{se}^\dagger a_{so} a_{se}^\dagger a_{io}^\dagger$. The first two correspond to actual processes. If they act on the vacuum state, the results show $a_{so}^\dagger a_{se} a_{se}^\dagger a_{io}^\dagger |0\rangle = (a_{so}^\dagger |0_{so}\rangle)(a_{se} a_{se}^\dagger |0_{se}\rangle)(a_{io}^\dagger |0_{io}\rangle) = |1_{so}, 0_{se}, 1_{io}\rangle$ and $a_{se}^\dagger a_{so} a_{so}^\dagger a_{ie}^\dagger |0\rangle = (a_{se}^\dagger |0_{se}\rangle)(a_{so} a_{so}^\dagger |0_{so}\rangle)(a_{ie}^\dagger |0_{ie}\rangle) = |1_{se}, 0_{so}, 1_{ie}\rangle$. For the last two terms, they don't affect the final entangled state because $a_{se}|0\rangle = 0$ and $a_{so}|0\rangle = 0$. Therefore the parallel-polarization entangled state is written as

$$|\Psi\rangle = |0\rangle + \int dv \int dv' P_{oo} h(L\Delta\beta_{eo}) h(L\Delta\beta_{oeo}) a_{so}^\dagger a_{io}^\dagger |0\rangle$$
$$+ \int dv \int dv' P_{ee} h(L\Delta\beta_{oe}) h(L\Delta\beta_{ooe}) a_{se}^\dagger a_{ie}^\dagger |0\rangle \quad (9)$$

with

$$P_{oo} = \frac{\pi^2 \varepsilon_0^2}{2} d_{31} \gamma_{51} \Omega_s E_{po} E_a G_{1,1} G_{3,-1} F_{EO} F_{eo} L^2 \sqrt{\frac{n_{so}^2 \Omega_s \Omega_i}{n_{io}^2 N_{se}^2 N_{io} N_{so}}}$$

$$P_{ee} = \frac{\pi^2 \varepsilon_0^2}{2} d_{31} \gamma_{51} \Omega_s E_{po} E_a G_{1,1} G_{3,1} F_{EO} F_{oe} L^2 \sqrt{\frac{n_{se}^2 \Omega_s \Omega_i}{n_{ie}^2 N_{so}^2 N_{ie} N_{se}}}$$

Similarly, we also set $\omega_s = \Omega_s + v$ and $\omega_i = \Omega_i - v$, while $\omega_s$ ($\omega_i$) is substituted by $\Omega_s$ ($\Omega_i$) in $P_{oo}$ and $P_{ee}$. Summarizing the two situations above, we rewrite the entangled state vector in a terser formation:

$$|\Psi\rangle = |0\rangle + (\int_{0-}^{0+} \delta(E_a) dE_a)[\int dv P_{oe} h(L\Delta\beta_{ooe}) a_{so}^\dagger a_{ie}^\dagger |0\rangle$$
$$+ \int dv P_{eo} h(L\Delta\beta_{oeo}) a_{se}^\dagger a_{io}^\dagger |0\rangle] + (1 - \int_{0-}^{0+} \delta(E_a) dE_a)$$
$$\times [\int dv \int dv' P_{oo} h(L\Delta\beta_{eo}) h(L\Delta\beta_{oeo}) a_{so}^\dagger a_{io}^\dagger |0\rangle \quad (10)$$
$$+ \int dv \int dv' P_{ee} h(L\Delta\beta_{oe}) h(L\Delta\beta_{ooe}) a_{se}^\dagger a_{ie}^\dagger |0\rangle]$$

When $E_a = 0$, we have $\int_{0-}^{0+} \delta(E_a) dE_a = 1$. The generated entangled state corresponds to $C_{oe}|o\rangle_s|e\rangle_i + C_{eo}|e\rangle_s|o\rangle_i$. If $E_a$ is an appropriate nonzero value, the value of $1 - \int_{0-}^{0+} \delta(E_a) dE_a$ should be 1, the entangled state vector is $C_{oo}|o\rangle_s|o\rangle_i + C_{ee}|e\rangle_s|e\rangle_i$. To ensure the EO transformation of the downconverted photons, several calculations have been done. The crystal length is set at 3 cm, and the average pump power is set at 100 W. Based on the coupled wave equations for EO interaction [6] and those for downconversion process, we obtain that when $E_a$ is $4.5 \times 10^5$ V/m, the transformation efficiencies are $P_{se}/(P_{se} + P_{so}) = 0.9958$ (for $|o\rangle_s|e\rangle_i \rightarrow |e\rangle_s|e\rangle_i$) and $P_{so}/(P_{se} + P_{so}) = 0.9971$ (for $|e\rangle_s|o\rangle_i \rightarrow |o\rangle_s|o\rangle_i$), respectively. The orthogonally polarized entangled pair could be transformed into parallel polarized entangled pair with a near 100% efficiency. Therefore, EO tunable entangled states are realized.

## 4. CHARACTERIZATION OF THE TUNABLE ENTANGLED SOURCE

To ensure the quality of our tunable entangle source, we investigate the corresponding properties. Firstly, the spectrum character is discussed, which is represented by the modulus squares of h-functions. For orthogonal polarization entangled photons, the corresponding expressions are $|h(L\Delta\beta_{ooe})|^2$ and $|h(L\Delta\beta_{oeo})|^2$; while those for the parallel-polarization entangled photons are simultaneously affected by SPDC process and EO interaction, so the formulations are $|h(L\Delta\beta_{oe})h(L\Delta\beta_{ooe})|^2$ and $|h(L\Delta\beta_{eo})h(L\Delta\beta_{oeo})|^2$, respectively. Calculation results are presented in Fig. 3. The natural spectral bandwidth of entangled source is mainly influenced by group velocities of the signal and idler wavelengths. In our situation, $\lambda_s$ and $\lambda_i$ are close with each other, so the differences between the bandwidths for two SPDC processes $o_p \rightarrow o_s + e_i$ and $o_p \rightarrow e_s + o_i$ are relatively small. From Fig. 3(a), if the waveguide length L is 5 cm, these two quantities are 0.21 nm and 0.17 nm, respectively. Compared with the previous report [11], these values are smaller, which is beneficial to the improvement of entanglement degree. For $C_{oo}|o\rangle_s|o\rangle_i + C_{ee}|e\rangle_s|e\rangle_i$, as EO process is activate, the natural bandwidths are further restricted. If L is 5 cm, the bandwidths for ordinary and extraordinary photon pairs are 0.13 nm and 0.15 nm, respectively, which are also adjustable in changing the PPLN's parameters.

The anticorrelation dip could be calculated based on the spectrum function, which is usually different for type-I and type-II SPDC [14]. The corresponding formulation is $R_C(\tau) \sim 1 - (1/2\pi) \int |h(v)|^2 \cos(v\tau) dv$. For the orthogonally polarized entangled pair, $|h(v)|^2$ corresponding to $\text{sinc}^2(L\Delta\beta_{ooe}/2)$ and $\text{sinc}^2(L\Delta\beta_{oeo}/2)$. We expand the phase mismatching to the first nonzero order of v. The corresponding formulations are $\Delta\beta_{ooe} \approx -(\beta'_{so} - \beta'_{ie})v$ and $\Delta\beta_{oeo} \approx -(\beta'_{se} - \beta'_{io})v$. $\beta'_{k\sigma}$ (k = p, s, i; σ = o,

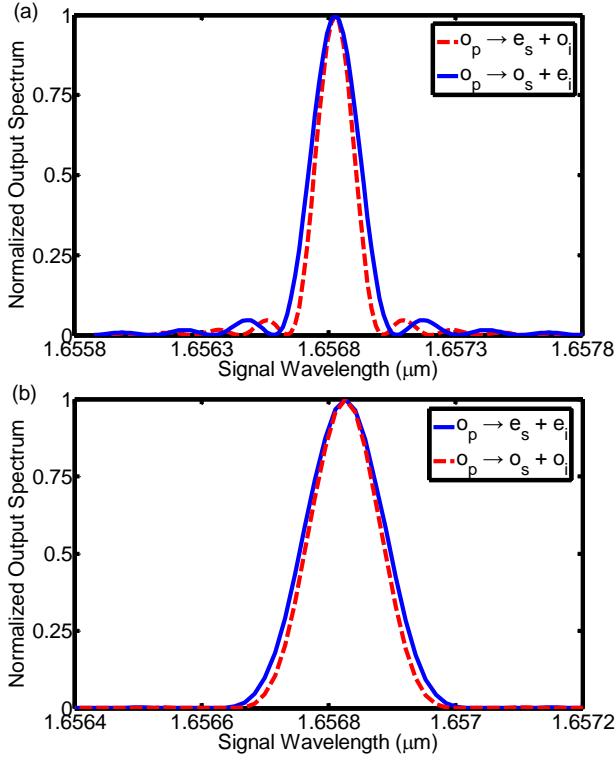

Fig. 3. Normalized output signal spectra corresponding to the two type of entangled states, respectively. (a) $o_p \to o_s + e_i$ and $o_p \to e_s + o_i$; (b) $o_p \to o_s + e_i \to e_s + e_i$ and $o_p \to e_s + o_i \to o_s + o_i$. The corresponding natural bandwidths are 0.17 nm, 0.21 nm, 0.13 nm and 0.15 nm, when the waveguide length L is 5 cm.

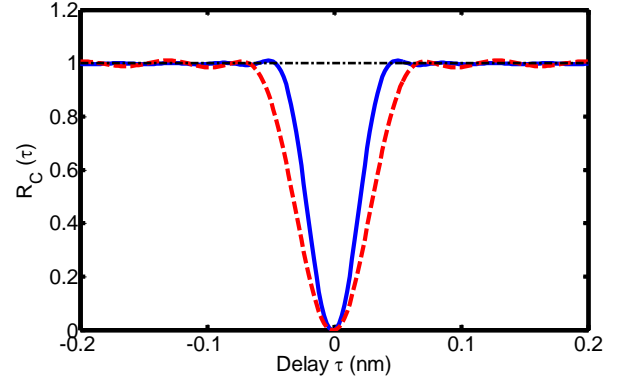

Fig. 4. The anticorrelation dips $R_C(\tau)$ are calculated based on $\text{sinc}^2(L\Delta\beta_{ooe}/2)$ and $\text{sinc}^2(L\Delta\beta_{ooe}/2) \text{sinc}^2(L\Delta\beta_{oe}/2)$, respectively. The blue solid line corresponds to the former, while the red dashed line represents the latter. The crystal length is set at 5 cm.

e) represents the dispersion parameter. For the parallel polarized entangled pair, as it is generated through EO transformation, the spectrum property is naturally different from the traditional cases. The $|h(\nu)|^2$ functions are written as $\text{sinc}^2(L\Delta\beta_{ooe}/2) \cdot \text{sinc}^2(L\Delta\beta_{oe}/2)$ and $\text{sinc}^2(L\Delta\beta_{oeo}/2) \cdot \text{sinc}^2(L\Delta\beta_{eo}/2)$. Similarly, we have $\Delta\beta_{oe} \approx (\beta'_{so} - \beta'_{se})\nu$ and $\Delta\beta_{eo} \approx (\beta'_{se} - \beta'_{so})\nu$. As an illustration, the anticorrelation dips corresponding to $\text{sinc}^2(L\Delta\beta_{ooe}/2)$ and $\text{sinc}^2(L\Delta\beta_{ooe}/2) \cdot \text{sinc}^2(L\Delta\beta_{oe}/2)$ are plotted in Fig. 4.

Based on the discussions above, we calculate the von Neumann entropy of our source. It is defined as $S = -\text{tr}(\rho_{\text{sub}} \log_2 \rho_{\text{sub}})$ [15]. $\rho_{\text{sub}}$ represents the reduced density operator for the subsystems. For a disentangled product state, S vanishes. If the state is maximally entangled, the value of S is 1. For our tunable entangled source, we start with density operator of the entangled state, which is expressed as $\widehat{\rho_{si}} = (\int_{0_-}^{0_+} \delta(E_a) dE_a)|\varphi_O\rangle\langle\varphi_O| + (1 - \int_{0_-}^{0_+} \delta(E_a) dE_a)|\phi_P\rangle\langle\phi_P|$ with $|\varphi_O\rangle$ and $|\phi_P\rangle$ corresponding to Eq. (7) and Eq. (9), respectively. The reduced density operator could be obtained through $\widehat{\rho_s} = -\text{tr}_i(\widehat{\rho_{si}})$. Therefore S is derived as

$$S = -(\int_{0_-}^{0_+} \delta(E_a) dE_a)[(\lambda_{oe}/\lambda_O)\log_2(\lambda_{oe}/\lambda_O) \\ + (\lambda_{eo}/\lambda_O)\log_2(\lambda_{eo}/\lambda_O)] - (1 - \int_{0_-}^{0_+} \delta(E_a) dE_a) \quad (11) \\ \times[(\lambda_{oo}/\lambda_P)\log_2(\lambda_{oo}/\lambda_P) + (\lambda_{ee}/\lambda_P)\log_2(\lambda_{ee}/\lambda_P)]$$

where $\lambda_{oe} = |\int d\nu\, P_{oe} h(L\Delta\beta_{ooe})|^2$, $\lambda_{oe} = |\int d\nu\, P_{eo} h(L\Delta\beta_{oeo})|^2$, $\lambda_{oo} = |\int d\nu \int d\nu'\, P_{oo} h(L\Delta\beta_{eo}) h(L\Delta\beta_{oeo})|^2$ and $\lambda_{ee} = |\int d\nu \int d\nu'\, P_{ee} h(L\Delta\beta_{oe}) h(L\Delta\beta_{ooe})|^2$. Besides, $\lambda_O = \lambda_{oe} + \lambda_{eo}$ and $\lambda_P = \lambda_{oo} + \lambda_{ee}$.

The influences of natural bandwidth and parameter $P_{k\delta}$ (k, $\delta$ = o, e) are clearly reflected in the equations above. As is discussed above, the natural bandwidth is so small that we could simplify the calculation of S into the perfect phase matching situation. Thus S has the same formulation for two types of entangled states. This result verifies that EO process does not affect the entanglement degree. Besides, $P_{k\delta}$ is also critical parameter for S. From Eq. (10), we could see that $P_{k\delta}$ is mainly influenced by geometry and material parameters. We find that the source has high tolerance for the geometrical variations. Until the waveguide modes of $\lambda_s$ and $\lambda_i$ are nearly cutoff, S still stays above 0.99. The degree of entanglement keeps in a high level.

## 5. DISCUSSIONS AND CONCLUSIONS

Generating the entangled states and manipulating them through EO interaction, what we have demonstrated above is merely a simplest case for the two-photon entangled state. And, this type of system could be used to generate multi-photon entanglement. Previously proposed generation methods of these states are usually complex and imperfect [16]. If the EO processes are also introduced, the instantly switching function thus could be effectively integrated in the circuit; on the other hand, since the EO effect may exchange the polarization of entangled photons effectively without breaking their entanglement, multi-photon entangled state with various configurations even could be generated. Besides, the mixed state entanglement plays important roles in the relevant quantum error correction processes [17], and polarization manipulation is a powerful approach to regulate the corresponding state. Therefore the SPDC and EO function-integrated circuit is also a very promising solution. For other special entangled states, *i.e.*, the nonmaximally entangled state [18], they also could be easily realized with the help of artificially designed domain structure. We may design a domain structure of LN to satisfy the QPM conditions for EO and SPDC processes of the pump light simultaneously. The ratio of the e-polarization and o-polarization portion of the pump photon is determined by the applied voltage, so the amplitude of eigenvectors of the entangled state becomes tunable. It's worth mentioning that besides the uniform domain periods we used all above, the domain structure of LN or similar materials could be elaborately prepared in more complicated patterns, such as quasi-periodic structures, which could provide complex reciprocal vectors to compensate the multi-process phase mismatching with high efficiency [19]. We believe the domain engineered nonlinear waveguide is really a very promising platform for quantum integration circuits, which deserves more in-depth studies in both theories and experiments [20].

In summary, we propose an effective approach to generate and manipulate entangled photons through the combination of EO and SPDC effects. Based on the domain-engineered PPLN waveguide, an EO tunable polarization entangled source is realized. The phase matching conditions of two SPDC processes and an EO interaction are simultaneously satisfied. It could selectively generate orthogonal-polarization or parallel-polarization photon by adjusting the applied voltage. The bandwidths and entanglement degrees of both types of entangled states are of excellent performances. Moreover, some kinds of other entangled states are discussed based on our proposals, including multi-photon entangled state, mixed entangled state and nonmaximally entangled state, showing some unique features. The introducing of EO polarization tuning into domain engineered nonlinear waveguides really provides a potential platform for the integrated and multi-functional quantum circuits.

ACKNOWLEDGMENTS

The authors thank the technical support from Miss Ting-ting Xu. This work is sponsored by 973 programs with contract No. 2011CBA00200 and 2012CB921803, and the National Science Fund for Distinguished Young Scholars with contract No. 61225026. The authors also thank the supports from PAPD and Fundamental Research Funds for the Central Universities.